\documentclass[preprint2]{aastex62}
\newcommand{\speed}[1]{#1 km~s${}^{-1}$}
\newcommand{\accel}[1]{#1 km~s${}^{-2}$}
\newcommand{\nfig}[1]{Figure~\ref{#1}}

\received{...}
\revised{...}
\accepted{...}

\shorttitle{EUV Waves Driven by Expansions of Coronal Loops}
\shortauthors{Shen et al.}

\begin{document}

\title{EUV Waves Driven by the Sudden Expansion of Transequatorial Loops Caused by Coronal Jets}

\correspondingauthor{Yuandeng Shen}
\email{ydshen@ynao.ac.cn}

\author[0000-0001-9493-4418]{Yuandeng Shen}
\affiliation{Yunnan Observatories, Chinese Academy of Sciences,  Kunming, 650216, China}
\affiliation{State Key Laboratory of Space Weather, Chinese Academy of Sciences, Beijing 100190, China}
\affiliation{CAS Key Laboratory of Solar Activity, National Astronomical Observatories, Beijing 100012, China}
\affiliation{Center for Astronomical Mega-Science, Chinese Academy of Sciences, Beijing, 100012, China}
\author{Zehao Tang}
\affiliation{Yunnan Observatories, Chinese Academy of Sciences,  Kunming, 650216, China}
\affiliation{University of Chinese Academy of Sciences, Beijing 100049, China}
\author{Yuhu Miao}
\affiliation{Yunnan Observatories, Chinese Academy of Sciences,  Kunming, 650216, China}
\affiliation{University of Chinese Academy of Sciences, Beijing 100049, China}
\author{Jiangtao Su}
\affiliation{CAS Key Laboratory of Solar Activity, National Astronomical Observatories, Beijing 100012, China}
\affiliation{University of Chinese Academy of Sciences, Beijing 100049, China}
\author[0000-0002-7694-2454]{Yu Liu}
\affiliation{Yunnan Observatories, Chinese Academy of Sciences,  Kunming, 650216, China}
 
\begin{abstract}
We present two events to study the driving mechanism of extreme-ultraviolet (EUV) waves that are not associated with coronal mass ejections (CMEs), by using high resolution observations taken by the Atmospheric Imaging Assembly (AIA) on board {\em Solar Dynamics Observatory}. Observational results indicate that the observed EUV waves were accompanied by flares and coronal jets, but without CMEs that were regarded as the driver of most EUV waves in previous studies. In the first case, it is observed that a coronal jet ejected along a transequatorial loop system at a plane-of-the-sky (POS) speed of \speed{335 $\pm$ 22}, in the meantime, an arc-shaped EUV wave appeared on the eastern side of the loop system. In addition, the EUV wave further interacted with another interconnecting loop system and launched a fast propagating (QFP) magnetosonic wave along the loop system, which had a period of 200 s and a speed of \speed{388 $\pm$ 65}, respectively. In the second case, we also observed a coronal jet ejected at a POS speed of \speed{282 $\pm$ 44} along a transequatorial loop system and the generation of bright EUV wave on the eastern side of the loop system. Based on the observational results, we propose that the observed EUV waves on the eastern side of the transequatorial loop systems are fast-mode magnetosonic waves, and they were driven by the sudden lateral expansion of the transequatorial loop systems due to the direct impingement of the associated coronal jets, while the QFP wave in the fist case formed due to the dispersive evolution of the disturbance caused by the interaction between the EUV wave and the interconnecting coronal loops. It is noted that EUV waves driven by sudden loop expansions have shorter lifetimes than those driven by CMEs.
\end{abstract}


\section{Introduction}
Magnetohydrodynamics (MHD) waves in the solar atmosphere have been intensively studied for many years, mainly due to their importance for coronal heating and their applications in coronal seismology \citep{1947MNRAS.107..211A,1999Sci...285..862N,2001A&A...372L..53N,2005LRSP....2....3N,2007Sci...317.1192T,2011Natur.475..477M,2017SoPh..292....7L}. In the past several decades, various types of MHD waves have been detected in the solar atmosphere. For example, large-scale chromospheric Moreton waves \citep[e.g.,][]{1960AJ.....65U.494M,2010ApJ...708.1639M,2012ApJ...752L..23S,2018MNRAS.474..770K} and coronal Extreme Ultraviolet (EUV) waves \citep[e.g.,][]{1998GeoRL..25.2465T,2010ApJ...723L..53L,2011ApJ...732L..20C,2013ApJ...764...70Z,2013AA...556A.152X,2013ApJ...775...39Y,2017ApJ...834L..15Z}, slow \citep[e.g.,][]{1997ApJ...491L.111O} and fast modes of magnetosonic waves \citep[e.g.,][]{2011ApJ...736L..13L,2012ApJ...753...53S,2013AA...554A.144Y,2013SoPh..288..585S,2018ApJ...853....1S,2018MNRAS.477L...6S}, Alfv\'{e}n waves \citep[e.g.,][]{2007Sci...318.1580C,2009Sci...323.1582J}, and various types of oscillations and waves in coronal loops and filaments \citep[e.g.,][]{2014ApJ...795..130S,2014ApJ...786..151S,2016SoPh..291.3303P,2017ApJ...842...99L,2017ApJ...851...47Z,2018AA...611A..47Z,2018JASTP.172...40K}. For more information on solar MHD waves, one can refer to recent reviews on this topic \citep[e.g.,][]{2005LRSP....2....3N,2014SoPh..289.3233L,2015LRSP...12....3W}. 

Large-scale EUV waves in corona have attracted a lots of attentions since their discovery \citep{1997SoPh..175..571M,1998GeoRL..25.2465T}. In the past twenty years, a large number of studies have been performed to study the driving mechanism and the physical nature of EUV waves, because interpretations about these basic problems are controversial. It is unclear that the observed EUV waves are driven by flare pressure pulses or coronal mass ejections (CMEs), because of that EUV waves exhibit both wave and non-wave characteristics. Since the launch of the {\em Solar Dynamics Observatory} \citep[{\em SDO};][]{2012SoPh..275....3P}, the high temporal and high spatial resolution multi-wavelengths observations taken by the Atmospheric Imaging Assembly \citep[AIA;][]{2012SoPh..275...17L} on board the {\em SDO} provides us an unprecedented opportunity to clarify these long-standing but basic problems. Based on AIA data, more and more observational studies tend to support that EUV waves are fast-mode magneticsonic waves in nature and they are driven by the associated CMEs \citep[e.g.,][]{2002ApJ...572L..99C,2006ApJ...641L.153C,2011ApJ...732L..20C,2012ApJ...752L..23S,2017ApJ...851..101S,2017SoPh..292....7L,2013AA...556A.152X}. The wave nature of EUV waves are confirmed by a lots of wave phenomena such as their fast-mode speed, reflection, transmission, and refraction effects during their interaction with other magnetic structures \citep[e.g.,][]{2008ApJ...680L..81L,2008ApJ...685..629G,2009ApJ...691L.123G,2012ApJ...746...13L,2012ApJ...756..143O,2013ApJ...773L..33S,2013ApJ...775...39Y}. In particularly, \cite{2006ApJ...641L.153C} performed a statistical study to investigate the driver of EUV waves, and the authors found that EUV waves only appear when CMEs are present. In addition, non-CME-associated energetic flares do not cause  EUV waves. This study highly indicates that EUV waves are driven by CMEs rather than flare pressure pluses. In spite of this, there were a small number of reported EUV waves that were not associated with CMEs \citep[e.g.,][]{1999SoPh..190..467W,2000SoPh..193..161T,2013ApJ...776...58N}. In addition, a large number of theoretical works are also performed to understand the physical nature of EUV waves \citep[e.g.,][]{2001JGR...10625089W,2002ApJ...572L..99C,2002ApJ...574..440O,2009ApJ...700.1716W,2012SCPMA..55.1316M,2011ApJ...728....2D,2012ApJ...750..134D,2016SoPh..291...89V,2018MNRAS.474..770K}.

Some studies based on high temporal and high spatial resolution data have revealed the detailed evolution process and the relationship between EUV waves and CMEs. By using stereoscopic observations from two viewpoints and three-dimensional geometrical modeling of the CME and wave structures, \cite{2009ApJ...700L.182P} found that the EUV wave occupies and affects a much bigger volume than the associated CME. Therefore, the authors suggested that the observed EUV wave and the associated CME were separated in space, and the former is likely driven by the latter \citep[see also,][]{2011ApJ...734...84L}. \cite{2012ApJ...745L...5C} identified the separation of an EUV wave from the expanding CME bubble in a powerful active region {\em GOES} M2.5 flare, which provided the direct evidence for supporting the scenario that EUV waves are driven by expanding CMEs. Very recently, \cite{2017ApJ...851..101S} also observed similar separation process of a small-scale EUV wave ahead of an expanding loop that was associated with a mini-filament eruption and a less energetic {\em GOES} B1.9 flare in the quiet-Sun region. Although no CME was associated with this miniature eruption, the expanding loop system can be regarded as the CME prototype in the low corona. This observation and previous studies on small-scale EUV waves together suggested the similarity between small- and large-scale EUV waves, and they also confirmed that EUV waves are driven by CMEs \citep{2012ApJ...753L..29Z,2013MNRAS.431.1359Z}.  

Coronal jets are ubiquitous in the solar atmosphere, they are frequently observed in coronal holes, active regions, and the quiet-Sun regions \citep[e.g.,][]{2008AA...478..907C,2011RAA....11.1229Y,2014PASJ...66S..12Y,2012RAA....12..573C,2015ApJ...815...71C}. Basically, coronal jets are thought to be caused by magnetic reconnection between emerging bipoles and their ambient open fields, therefore, many studies reported that coronal jets are tightly associated with magnetic flux emergences \citep[e.g.,][]{2004ApJ...610.1136L,2007Sci...318.1591S,1995Natur.375...42Y,2011ApJ...735L..43S,2015ApSS.359...44L,2016ApJ...833..150L}. In addition, magnetic flux cancellations are also important for triggering coronal jets \citep[e.g.,][]{2007AA...469..331J,2012ApJ...745..164S,2017ApJ...851...67S,2016ApJ...832L...7P,2017ApJ...844..131P}. Based on high resolution observations, many studies revealed that a large number of coronal jets are tightly in association with the eruptions of mini-filaments, and such kind of jets are dubbed as blowout jets \citep[e.g.,][]{2010ApJ...720..757M,2012ApJ...745..164S,2017ApJ...851...67S,2012NewA...17..732Y,2012RAA....12..300Y,2015ApJ...814L..13L,2015Natur.523..437S,2016ApJ...830...60H,2016ApJ...821..100S,2017ApJ...835...35H,2013RAA....13..253H,2017SoPh..292..152J,2017ApJ...844L..20Z,2017ApJ...842L..20L,2018MNRAS.476.1286J,2018ApSS.363...26L}. We note particularly that sometimes a normal coronal jet can cause a single CME in the field-of-views (FOVs) of coronagraphs, and the shape of the associated CME often resemble the shape of the jet \citep[e.g.,][]{2002ApJ...575..542W,2005ApJ...628.1056L,2008SoPh..249...75L}. However, a blowout jet can cause a jet-like and a bubble-like CMEs simultaneously, in which the jet-like CME is resulted from the reconnection between open magnetic fields and the adjacent closed loops, while the bubble-like CME is caused by the eruption of the mini-filament confined in the jet-base \citep{2012ApJ...745..164S}. Recently, it is interesting that some observational studies reported that coronal jets are dynamically related to large-scale EUV waves. However, most of the jet-related EUV waves  are also provided to be driven by the associated CMEs \citep[e.g.,][]{2012ApJ...747...67Z,2013ApJ...764...70Z}. In particularly, \cite{2015ApJ...804...88S} reported an EUV wave that was in association with a coronal jet and a type \uppercase\expandafter{\romannumeral2} radio burst, but the event did not accompanied by a CME. The authors identified that the type \uppercase\expandafter{\romannumeral2} radio burst was produced by the EUV wave, and they further proposed that the generation of the EUV was possibly caused by the expansion of the newly formed magnetic loops during the magnetic reconnection  that produced the observed coronal jet.

\begin{figure*}[thbp]
\epsscale{1}
\figurenum{1}
\plotone{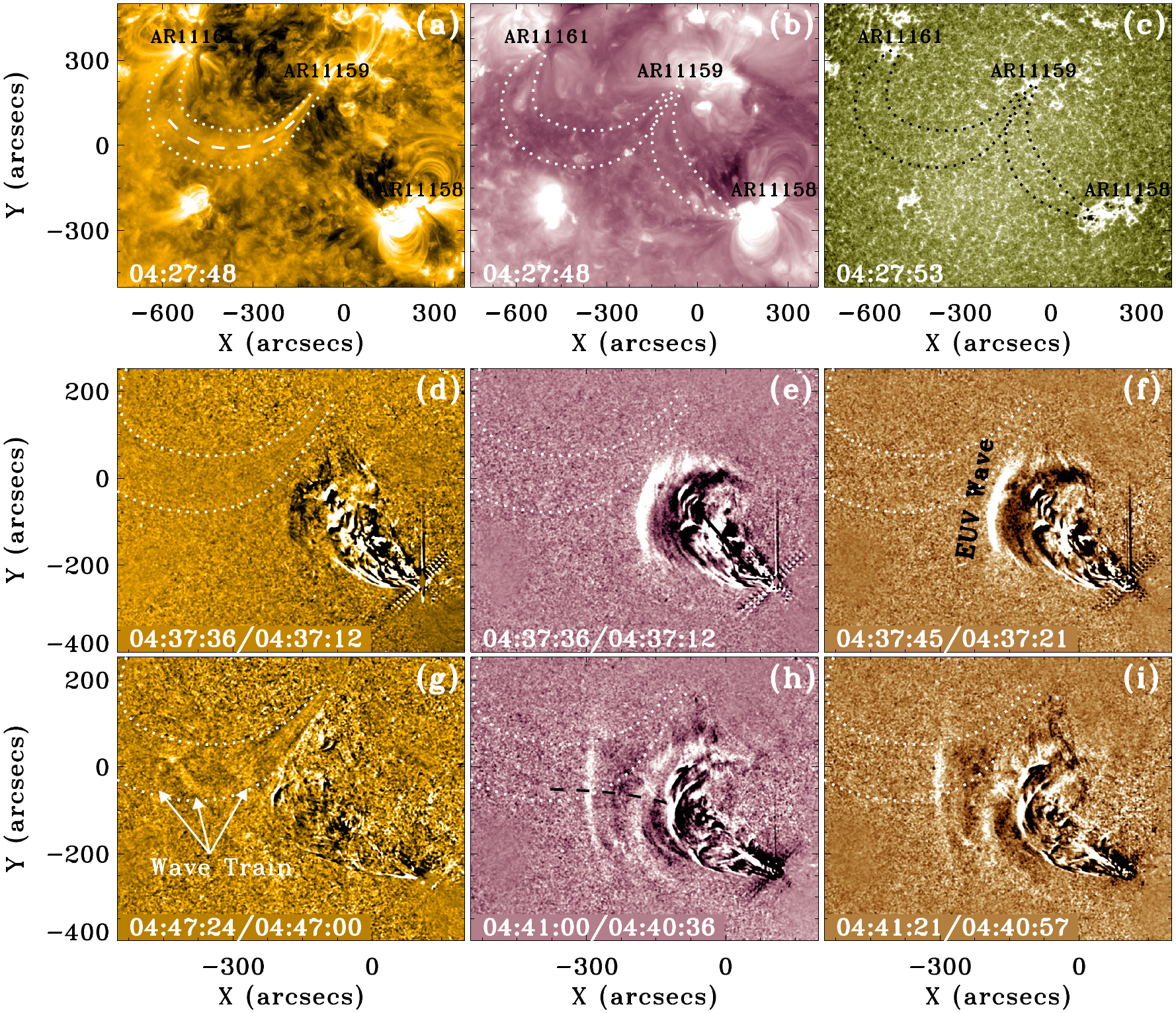}
\caption{The eruption of the event on 2011 February 15. The top row shows the AIA 171 (a), 211 (b), and 1600 \AA\ (c) direct images, while the left, middle, and right columns of the bottom two rows display the running ratio images of AIA 171, 211, and 193 \AA\, respectively. The loop system connecting AR11161 and AR11159 is outlined with two dotted curves in panel (a), while the one connection AR11158 and AR11159 is also outlined in panel (b). The contour of the loop connecting AR11161 and AR11159 is also overlaid in other panels. The three arrows in panel (g) indicate the wave fronts in the closed loop, while the EUV wave ahead of the jet is annotated in panel (f). The dashed curves in panel (a) and (h) indicate the paths that are used to obtain the time-distance diagrams of the waves. An animation for this figure is available in the online journal.
\label{fig1}}
\end{figure*}

Generally speaking, any disturbance can cause a wave in the coronal plasma. For magnetosonic waves, both the magnetic and the plasma pressure act as the restoring forces. Considering the complicated magnetic structures in the solar corona, the generation of fast-mode EUV waves should have other drivers besides CMEs. In this paper, we observed that large-scale EUV waves can be launched by the lateral expansion of coronal loops caused by the direct impingement of coronal jets. Two similar events are presented in this paper to illustrate this new driving mechanism of large-scale EUV waves. Main observational results are described in Section 2; interpretation of the driving mechanism of the EUV waves are presented in Section 3; Conclusion and discussion are given in the last section.

\begin{figure*}[thbp]
\epsscale{1}
\figurenum{2}
\plotone{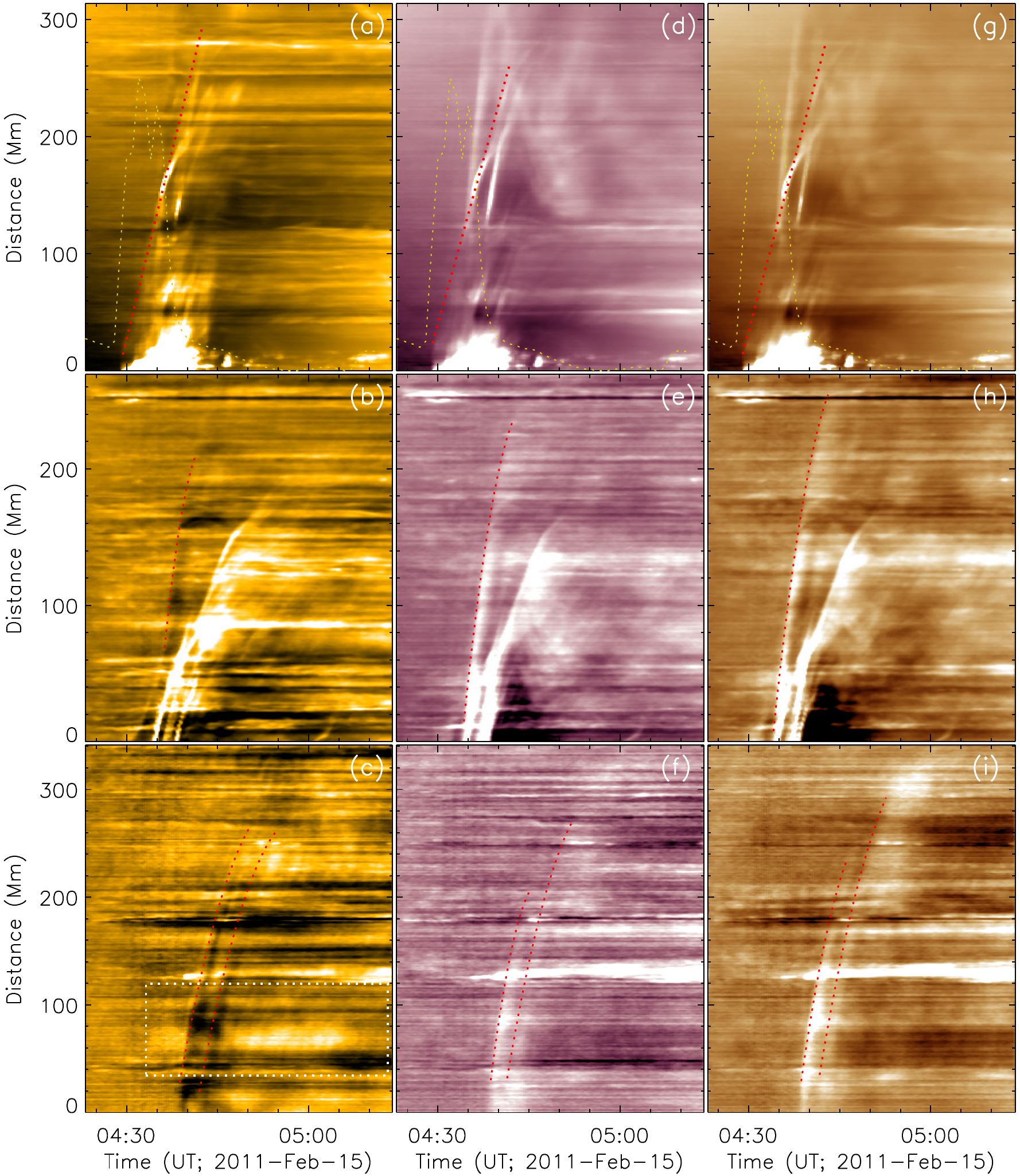}
\caption{Time-distance diagrams show the kinematics of the jet (top row), the EUV wave ahead of the jet (middle row), and the wave fronts along the closed loop connecting AR11161 and AR11159 (bottom row). The left, middle, and right columns are time-distance diagrams made from the AIA 171, 211, and 193 \AA\ running ratio images, respectively. The yellow dotted curves overlaid in the top row are the {\em GOES} soft X-ray flux in the energy band of 1 -- 8 \AA\, while the red dotted curves are the quadratic fit to the moving jet (top row) and the wave fronts (bottom two rows). The time-distance diagrams in the top row are made along the jet axis, while those in the middle and bottom rows are made along the dashed curves as shown in \nfig{fig1} (a) and (h), respectively.
\label{fig2}}
\end{figure*}

\section{Results}
\subsection{The Event on 2011 February 15}
On 2011 February 15, a {\em GOES} C8.3 flare occurred in NOAA active region AR11158. The start, peak, and stop times of the flare were 04:29:00, 04:49:00, and 05:09:00 UT, respectively. The flare was in association with a coronal jet, but it did not cause any detectable CME in coronagraphs. The AIA 171, 211, and 1600 \AA\ images are displayed in the top row of \nfig{fig1} to show the pre-eruption magnetic conditions. It can be seen that three active regions, AR11158, AR11159, and AR11161, were involved in this event, in which AR11158 located in the southern hemisphere, while the other two located in the northern hemisphere. As outlined by the white dotted curves, these active regions were connected by two groups of loop systems. The one connected AR11161 and AR11159, while the other is a transequatorial loop system that connected AR11158 and AR11159.

The eruption of the jet is displayed in the bottom two rows of \nfig{fig1} with running ratio images of AIA 171, 211, and 193 \AA\ images, and the dotted curves in each panel show the position of the loop system that connected AR11159 and AR11161. Here, a running ratio image is obtained by dividing an image by the one taken at 24 s before, in which moving features can be observed more clearly. It should be noted that only two moments of the eruption are shown in \nfig{fig1}, and one can see the online animation for more detailed information.  The coronal jet erupted at about 04:29:00 UT from AR11158 and along the transequatorial loop system. It is noted that the ejecting material of the jet was trapped in the transequatorial loop system. During the ejection, a bright wavefront appeared on the eastern side of the transequatorial loop system at about 04:35:00 UT (see \nfig{fig1} (e) and (f)). The arc-shaped wavefront propagated eastwardly at the first, and then its propagation direction changed to southeast. It is noted that the curvature of the wavefront become increasingly smaller. In the meantime, the brightness of the wavefront became increasingly weaker, and it finally disappeared at about 04:45:00 UT. The lifetime of the EUV wave was about 10 minutes, which is much shorter than the typically hour-long lifetime of EUV waves \citep{2014SoPh..289.3233L}. Based on these observational results, we think that the generation of the eastward EUV wave was possibly due to the sudden eastward expansion of the transequatorial loop system caused by the sudden impingement of the coronal jet upon the loop system.

It is interesting that the northern section of the wavefront interacted with the other loop system that connected AR11161 and AR11159 at 04:38:38 UT, which caused obvious brightening around the interaction position. A few minutes later, multiple wavefronts appeared and propagated along the loop system that connected AR11159 and AR11161. The new formed wavefronts can be best seen in the AIA 171 \AA\ running ratio images (see the arrows in \nfig{fig1} (g)), and it could be regarded as a quasi-periodic fast propagating (QFP) wave \citep[e.g.,][]{2011ApJ...736L..13L,2012ApJ...753...53S,2013SoPh..288..585S,2018ApJ...853....1S,2018MNRAS.477L...6S}. In most of the previous studies, the generation of QFP waves were found to be associated with the pulsations in the accompanying flares. Here, based on the close temporal and spatial relationship between the QFP wave and the EUV wave, it is obvious that the generation of the multiple wavefronts were associated with the interaction of the EUV wave. Therefore, the generation of this QFP wave was possibly due to the dispersive evolution of the disturbance caused by the interaction \citep{2012ApJ...753...52L,2018MNRAS.477L...6S}. Here, we would like to point out that mode conversion of solar EUV wavefronts have been reported in recent observational and simulation studies, where fast-mode magnetosonic waves transformed into slow-mode magnetosonic waves during their interaction with other magnetic structures \citep{2016ApJ...822..106C,2016SoPh..291.3195C,2017ApJ...834L..15Z}.

\begin{figure*}[thbp]
\epsscale{1}
\figurenum{3}
\plotone{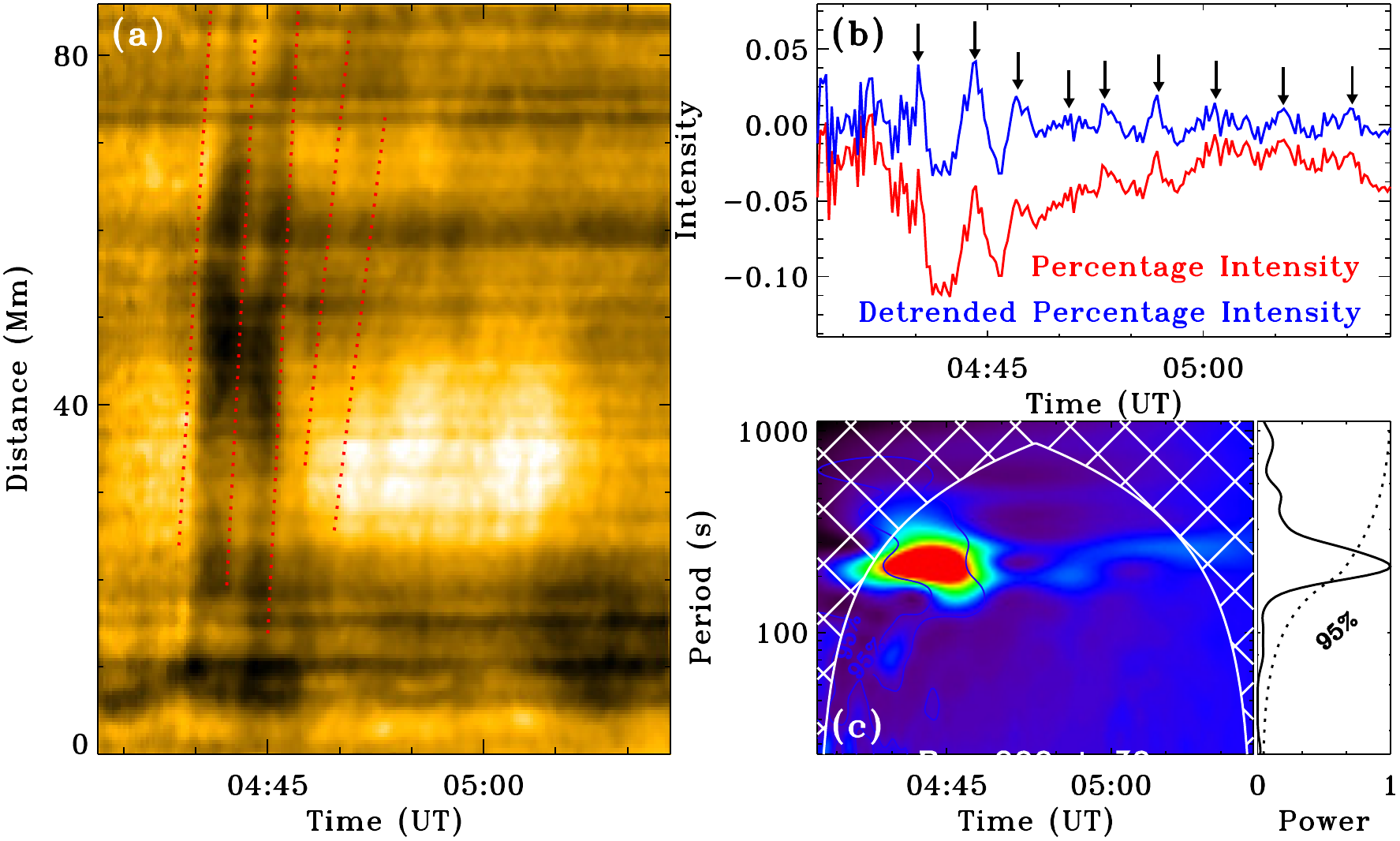}
\caption{Panel (a) is a partial AIA 171 \AA\ time-distance diagram taken from the white box region as shown in \nfig{fig2} (e), in which the red dotted lines indicate the propagating wavefronts. Panel (b) shown the percentage intensity (red) and its detrended (blue) profiles along the white dashed line in panel (a). The vertical arrows in panel (b) indicate the positions of the wavefronts. Panel (c) shows the wavelet power map of the intensity profile, and the global power is plotted on the right.
\label{fig3}}
\end{figure*}

The kinematics of the coronal jet and the waves are studied with time-distance diagrams. In a time-distance diagram, a moving feature can be observed as an inclined bright stripe whose slope represent the moving speed. To obtain a time-distance diagram, one need to first obtain the one-dimensional intensity profiles along a specified path at different times, then a two-dimensional time-distance diagram can be generated by stacking the obtained one-dimensional intensity profiles in time. The top row of \nfig{fig2} shows the time-distance diagrams made from AIA 171, 211, and 193 \AA\ running ratio images along the main axis of the jet body. By applying a linear fit to the bright stripe, it is obtained that the average speed of the jet is about \speed{335 $\pm$ 22}. The time-distance diagrams along the black dashed curve as shown in \nfig{fig1} (h) are plotted in the middle row of \nfig{fig2}, which show the kinematics of the EUV wavefront on the eastern side of the transequatorial loop system. It can be seen that the wavefront show obvious deceleration during its propagation. By fitting the stripes with a linear (quadratic) functions, it can be obtained that the mean speed (deceleration) of the wavefront was about \speed{466 $\pm$ 12} (\accel{-0.65 $\pm$ 0.23}). The bottom row of \nfig{fig2} shows the time-distance diagrams made along the white dashed curve as shown in \nfig{fig1} (a), which show the kinematics of the QFP wave along the loop system that connected AR11159 and AR11161. There are two strong stripes can be clearly identified in the time-distance diagrams, and the average speed (deceleration) of the wavefronts was about \speed{388 $\pm$ 65} (\accel{-0.38 $\pm$ 0.17}). It should be pointed out that the speeds of the jets and waves in the present paper are the POS speeds on the sky plane, which are the lower limits of their true three-dimensional values.

\begin{figure*}[thbp]
\epsscale{0.98}
\figurenum{4}
\plotone{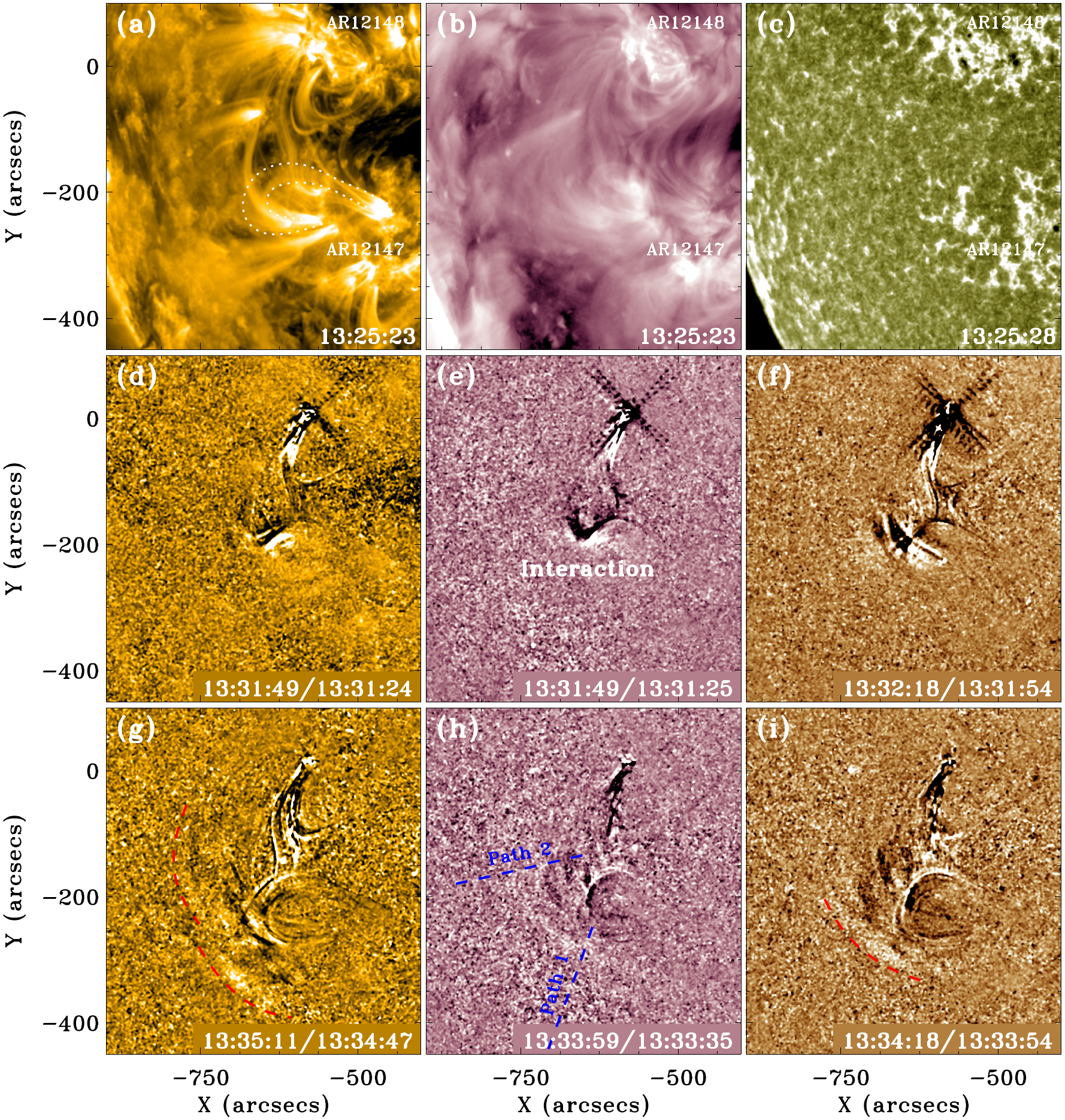}
\caption{The eruption of the event on 2014 August 21. The same with \nfig{fig1}, the top row show the AIA 171 (a), 211 (b), and 1600 \AA\ (c) images, while the left, middle, and right columns of the bottom two rows display the running ratio images of AIA 171, 211, and 193 \AA\, respectively. The dotted curves in panel (a) outline a closed loop in AR12147, and the active regions AR12148 and AR12147 are also indicated. The red dashed curves in panels (g) and (i) indicate the wave fronts appeared after the interaction, while the two blue dashed lines in panel (h) show the paths that are used to obtain the time-distance diagrams of the waves. An animation for this figure is available in the online journal.
\label{fig4}}
\end{figure*}

To further analyze the properties of the QFP wave, the details of the white box region is zoomed in in \nfig{fig3} (a), from which one can observe many stripes that represent the propagating QFP wavefronts, and the first five wavefronts are highlighted with red dotted lines. The percentage and its detrended intensity profiles at the position indicated by the white dashed line in \nfig{fig3} (a) are plotted in \nfig{fig3} (b), which show the wavefronts more clearly, and the corresponding wavefronts are indicated by the vertical arrows. The periodicity of the intensity profile is analyzed by using the wavelet software \citep{1998BAMS...79...61T}, which reveals a strong period of about 200 $\pm$ 30 s in the QFP wave (see \nfig{fig3} (c)). 

\subsection{The Event on 2014 August 21}
On 2014 August 21, a flare occurred in NOAA active region AR12148 in the northern hemisphere of the Sun, which was associated a coronal jet heading to the southern hemisphere, but it did not cause any CME in coronagraphs. It is unable to distinguish the class of the flare, because it was covered by a large {\em GOES} M3.4 flare near the eastern limb of the solar disk. By observing the start of the brightening in EUV observations, it can be determined that the start time of the flare was at about 13:28:00 UT. The pre-eruption magnetic condition is shown in the top row of \nfig{fig4} with AIA 171, 211, and 1600 \AA\ images. The active regions AR12147 and AR12148 were involved in this event, and they were connected by a group of transequatorial loop system. In addition, there is a closed loop system connected the opposite magnetic polarities in AR12147, and it was outlined by the two dotted curves in \nfig{fig4} (a).

\begin{figure*}[thbp]
\epsscale{1}
\figurenum{5}
\plotone{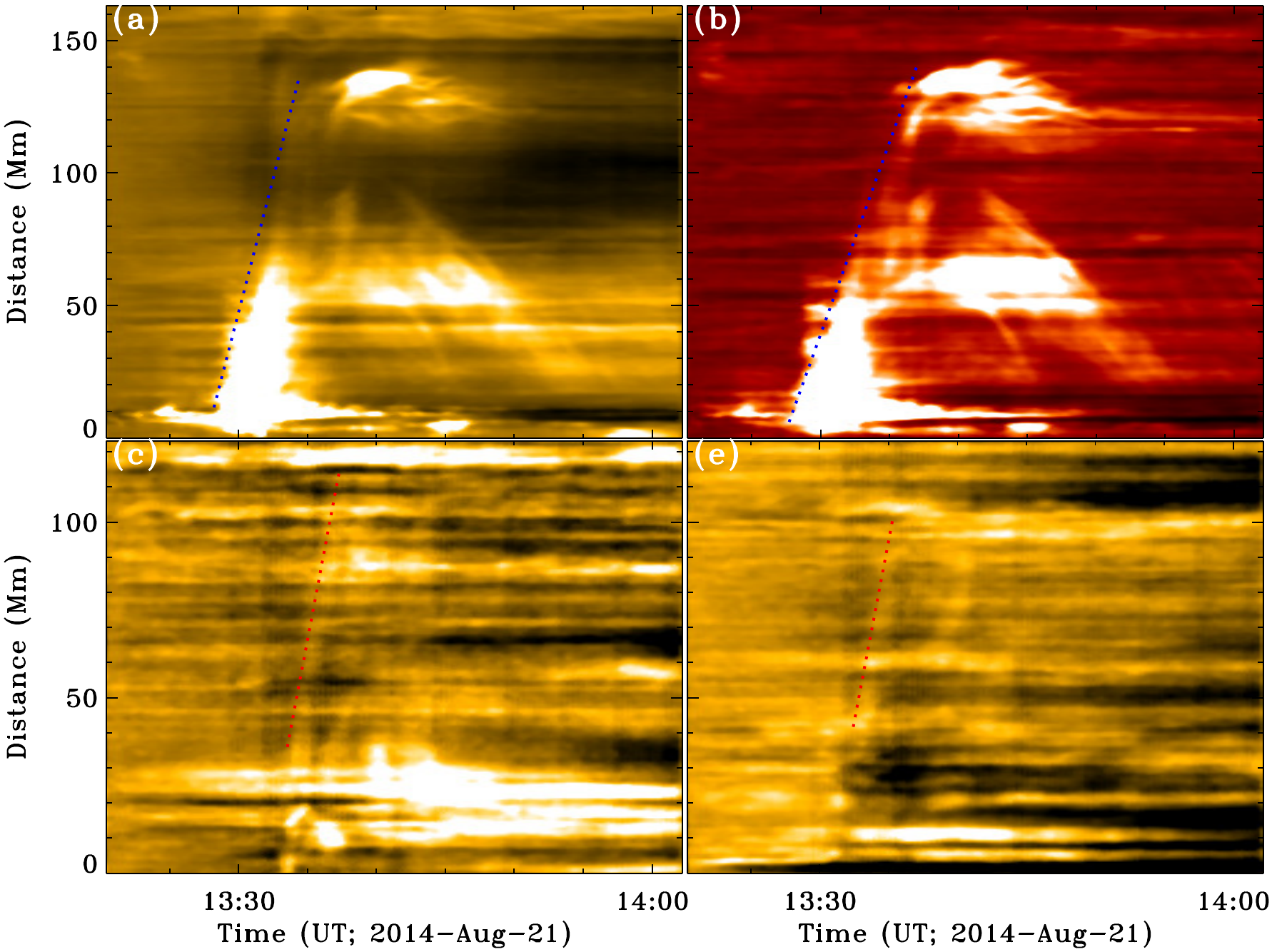}
\caption{Time-distance diagrams show the kinematics of the jet and the wave fronts. Panels (a) and (b) show the time-distance diagrams made from the AIA 171 and 304 \AA\ running ratio images along the jet axis, respectively. Panels (c) and (d) are time-distance diagrams made from AIA 171 \AA\ running ratio images along path 1 and 2 as shown in \nfig{fig3}, respectively. The blue and red dotted lines in the figure are the linear fit to the moving jet (top row) and the wave fronts (bottom row). 
\label{fig5}}
\end{figure*}

Simultaneously with the start of the flare, a coronal jet ejected outward from the flaring region in AR12148. The ejection of the coronal jet was in the south direction and along the transequatorial loops, which directly impinged on the closed loop system in AR12147 at around 13:31:49 UT, and caused obvious brightening around the interaction region and sudden movement of the loop system in the southeast direction (see the middle row of \nfig{fig4}). After the start of the interaction, a part of the ejecting plasma stopped around the interaction region, while the other part of the ejecting plasma moved back along the ejection path. This pattern of mass motion is similar to the failed filament eruptions in which the erupting filament plasma is prevented to erupt by the overlying magnetic fields \citep[e.g.,][]{2011RAA....11..594S,2012ApJ...750...12S}. In addition, the closed loop system showed oscillation motion after the interaction, due to the return motion of the loop system. More interestingly, the interaction between the jet and the closed loop system directly launched multiple bright wavefronts in the eastern side of the the closed loop system, which are highlighted with red dashed curves in \nfig{fig4} (g) and (i). It is noted that the wavefronts can be divided into two sets based on their propagation directions, in which one was in the east direction, while the other was in the southeast direction. Here, the eastward propagating wavefronts were probably excited by the eastward expansion of the transequatorial loops like the EUV wave observed in the first case, while the southeastward wavefronts were possibly caused by the sudden southeastward motion of the closed loop due to the interaction of the coronal jet.

The kinematics of the coronal jet and the wavefronts are studied with time-distance diagrams, and the results are shown in \nfig{fig5}. The time-distance diagrams made from the AIA 171 and 304 \AA\ running ratio images along the main axis of the coronal jet are plotted in \nfig{fig5} (a) and (b), respectively. It can be seen that the jet started at about 13:28:00 UT, then it ejected with a linear speed. It is measured that the speeds of the jet were about \speed{326 and 238} based on the AIA 171 and 304 \AA\ time-distance diagrams, respectively. Therefore, the average speed of the jet is \speed{282 $\pm$ 44}. When the coronal jet reached a distance of about 350 Mm from the eruption source region,  a part of the ejecting plasma stopped there for about 30 minutes due to the impediment of the closed loop system in AR12147, and the other part of the ejecting material started to move back along the transequatorial loop system. The time-distance diagrams plotted in \nfig{fig5} (c) and (d) are made along paths 1 and 2 as shown in \nfig{fig4} (h) that are along the propagation directions of the two sets of wavefronts. Since the wavefronts are very weak, in each time-distance diagram only one incline stripe can be clearly recognized. By applying a linear fit to the stripes, it is obtained that the propagation speeds of the waves along paths 1 and 2 were about \speed{320 and 360}, respectively. The appearance and disappearance times of the EUV waves were at about 13:32:47 UT and 13:39:59 UT, respectively. Therefore, the lifetimes of the EUV waves were about only 7 minutes.

\begin{figure*}
\epsscale{1}
\figurenum{6}
\plotone{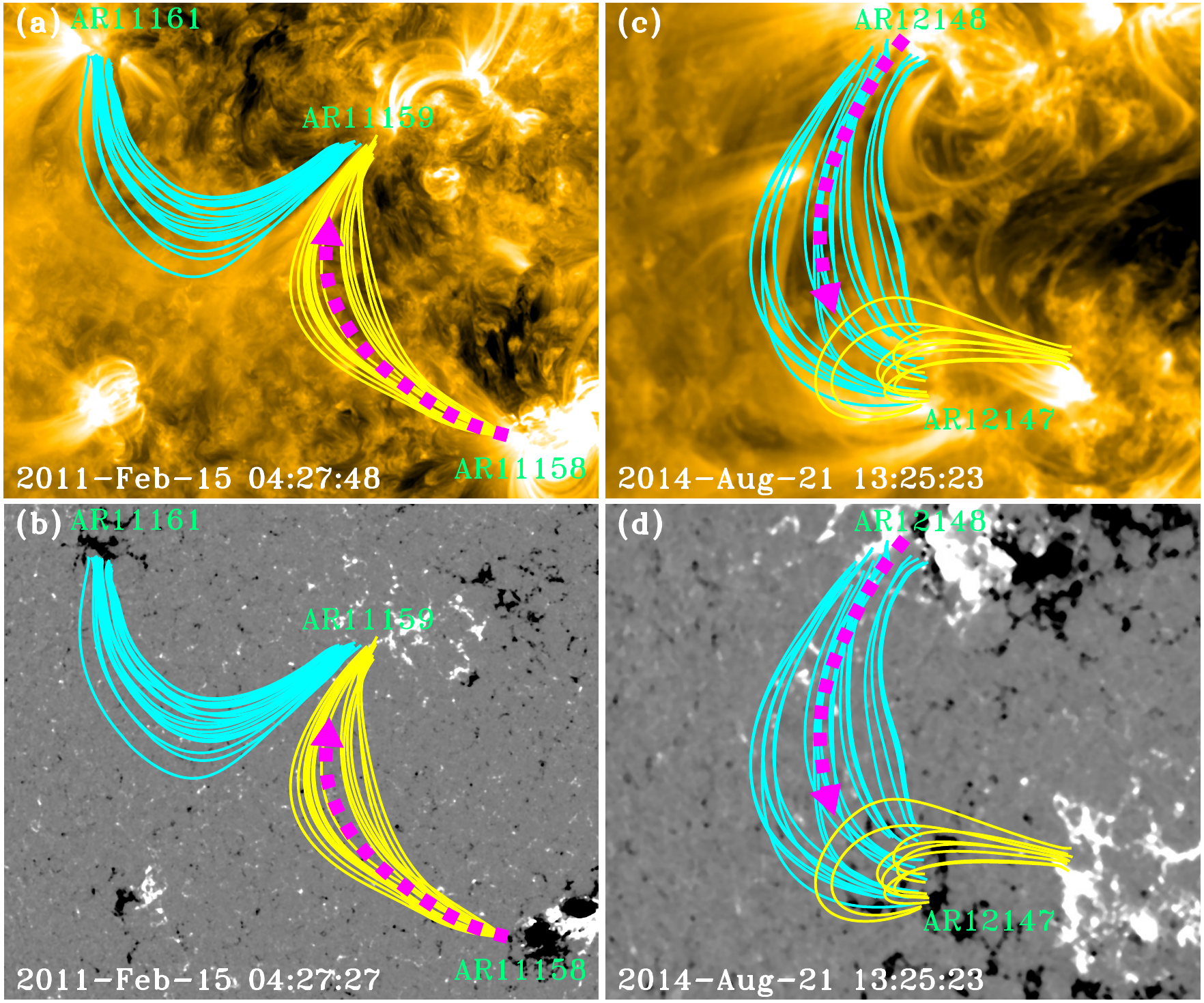}
\caption{Left and right columns show the AIA 171 \AA\ images (panels (a) and (c)) and the HMI line-of-sight magnetograms (panels (b) and (d)) on 2011 February 15 and 2014 August 21, respectively. In the magnetograms, the white and black patches represent the positive and negative magnetic polarities, respectively. The cyan and yellow curves are the extrapolated magnetic field lines based on the PFSS software. The dashed pink arrows indicate the ejection direction of the coronal jets. The numbers of the active regions are also plotted in the figure.
\label{fig6}}
\end{figure*}

\begin{figure*}[thbp]
\epsscale{1}
\figurenum{7}
\plotone{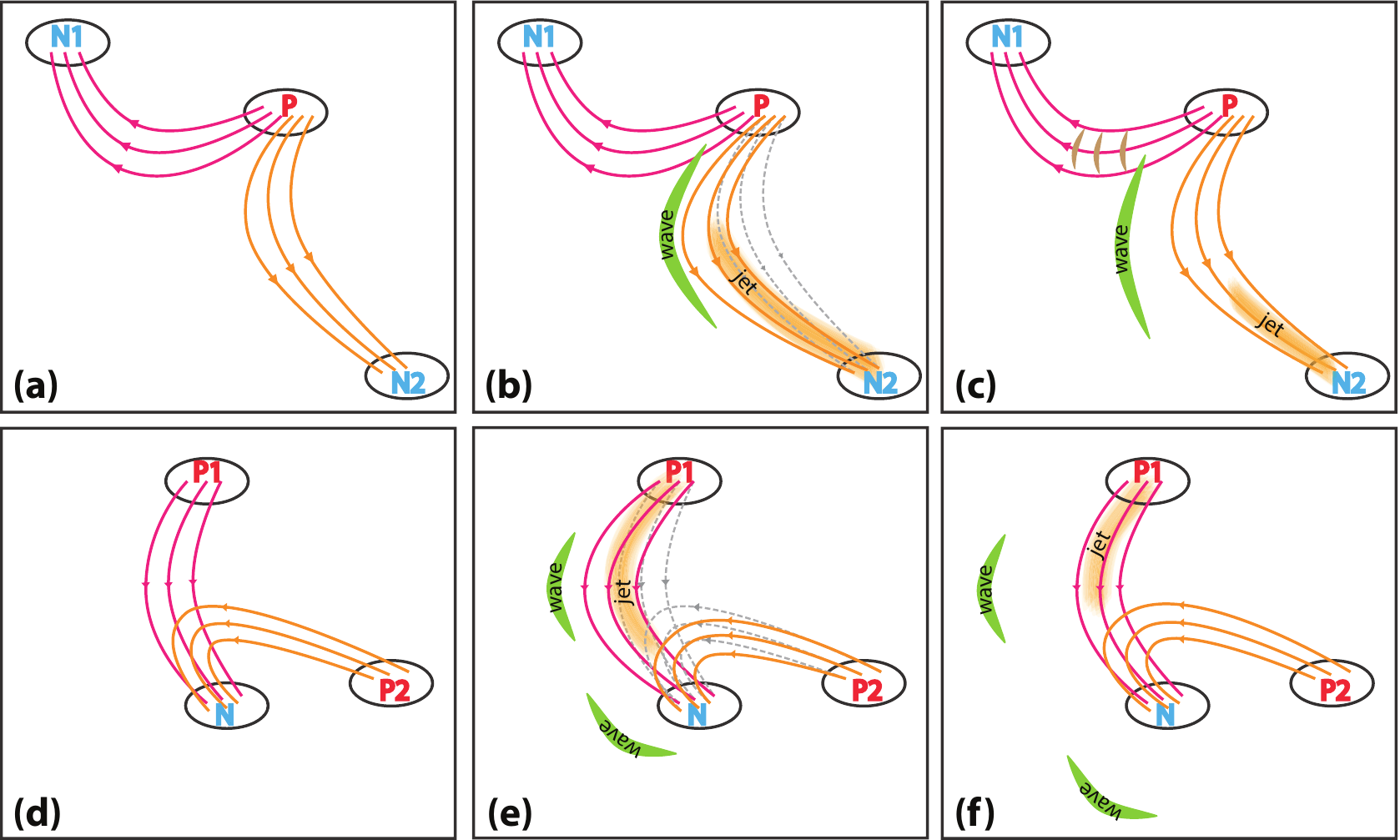}
\caption{Cartoon illustration of the driving mechanism of the EUV waves. The top and bottom rows are for the events on 2011 February 15 and 2014 August 21, respectively. The red and yellow curves are some representative field lines connecting the polarities. The dashed gray curves in panels (b) and (e) indicate the position of the undisturbed field lines. The wave fronts are plotted as green and brown colors, and the ejecting jets are also marked in the figure. The letters P, N1, and N2 in the top row represent the active regions AR11159, AR11161, and AR11158, respectively. In the bottom row, P1 represents the active region AR12148, while P2 and N indicate the positive and negative polarities of active regions AR12147, respectively.
\label{fig7}}
\end{figure*}

\subsection{Magnetic Field Extrapolation}
To better illustrate the magnetic topologies and the entire evolution processes of the two events, we further extrapolated the three-dimensional magnetic fields by using the potential field source surface (PFSS) software available in the SolarSoftWare (SSW) package \citep{2003SoPh..212..165S}. The extrapolated results of the events on 2011 February 15 and on 2014 August 21 are shown in the left and the right columns of \nfig{fig6}, respectively. It should be pointed out that only some representative magnetic field lines are overlaid on the AIA 171 \AA\ images and the LOS magnetograms before the eruptions. For the event on 2011 February 15, the loop system connected AR11161 and AR11158 is plotted in gray, the transequatorial loop system is plotted in yellow, and the yellow dashed arrow indicates the ejecting coronal jet. For the event on 2014 August 21, the transequatorial loop system is plotted in gray, while the closed loop system in AR12147 is plotted in yellow. In the same way, the ejecting coronal jet is indicated by the purple dashed arrow. It can be seen that the extrapolated magnetic field lines do not exactly match the actual coronal loops observed in the AIA 171 \AA\ images, which is due to the approximations used in the  PFSS extrapolation model in which the magnetic field is assumed to be potential. However, the extrapolated results do not affect the understanding of the large-scale magnetic topologies and the connectivities.

\section{Interpretation}
To better understand the driving mechanism of the observed EUV and QFP waves in the present study, we plot a cartoon in \nfig{fig7} to illustrate the detailed evolution processes of the two events. It should be noted that only some representative magnetic field lines are plotted in the figure, and the active regions are indicated as letters. The pre-eruption magnetic configuration of the event on 2011 February 15 is plotted in \nfig{fig7} (a), in which the letters P, N1, and N2 represent the active regions AR11159, AR11161, and AR11158, respectively. The ejection of the coronal jet from N2 can result in the sudden lateral expansion or movement of the transequatorial loop system in the east direction. This will abruptly increase the magnetic and gas pressure forces perpendicular to the transequatorial loop system, which therefore results in the generation of the observed EUV wave (see \nfig{fig7} (b)). The generation of the QFP wave is shown in \nfig{fig7} (c). Due to the eastward propagation of the EUV wave, it will interact with the loop system that connects P and N1. This interaction will result in a strong disturbance in the loop, which can further evolve into multiple wavefronts due to the dispersive evolution of the disturbance in the inhomogeneous plasma medium \citep{2018MNRAS.477L...6S}. 

The pre-eruption magnetic configuration of the event on 2014 August 21 is plotted in \nfig{fig7} (d), in which the letters P1, P2, and N represent the active region AR12148, and the positive and negative polarities of AR12147, respectively. Here, the generation of the EUV wave on the eastern side of the transequatorial loop system that connects P1 and N is the same with the one observed in the event on 2011 February 15 (see \nfig{fig7} (e) and (f)), while the generation of the EUV wave in the southeast direction is possibly caused by the sudden southeastward motion of the closed loop system that connects N and P2, due to the impingement of the coronal jet.

\section{Conclusions and Discussions}
Using high temporal and high spatial resolution observations taken by the {\em SDO}, two EUV wave events are analyzed in detail to study the generation mechanism of EUV waves. It is found that the EUV waves were in association with the violent eruption of coronal jets along transequatorial loops. 

For the event on 2011 February 15, the coronal jet ejected at a speed of about \speed{335 $\pm$ 22} along the transequatorial loop system that connected AR11158 and AR11159. During the ejecting of the coronal jet, a bright arc-shaped EUV wave appeared on the eastern side of the transequatorial loop system, which propagated outward in the east direction at an average speed (deceleration) of about \speed{466 $\pm$ 12} (\accel{-0.65 $\pm$ 0.23}). During the propagation of the EUV wave, the northern section of the EUV wavefront interacted with an other loop system that connected AR11159 and AR11161. The interaction between the wave and the loop system directly caused the generation of a QFP wave along the loop system. It is measured that the period of the QFP wave was about 200 s, while the average speed (deceleration) was about \speed{388 $\pm$ 65} (\accel{-0.38 $\pm$ 0.17}).

For the event on 2014 August 21, it is observed that a coronal jet ejected from AR12148 at an average speed of about \speed{282 $\pm$ 44}. The ejection of the coronal jet was along the transequatorial loop system that connected AR12147 and AR12148. The jet directly impinged on the closed loop system in AR12147, which caused the sudden southeastward movement of the loop system. After the interaction, the closed loop system showed obvious oscillation motion. In the meantime, EUV wavefronts are simultaneously observed in the southeast direction of the closed loop system and the eastern side of the transequatorial loop system. It is measured that the speeds of the EUV waves on the eastern side of the transequatorial loop system and the southeast direction of the closed loop system were about \speed{360 and 320}, respectively. The EUV waves observed in the first event showed obvious deceleration during their propagation, but those in the second case did not. In addition, the propagation speeds of the observed EUV waves in the two events are all significantly higher than the sound speed \citep[\speed{150 -- 210};][]{2014SoPh..289.3233L} and fall in the range of fast-mode speed \citep[\speed{230 -- 1500};][]{2007ApJ...664..556W} in the quiet-Sun corona. In addition, due to the projection effect the true wave speed should be larger than those we obtained from the imaging observations. Therefore, these EUV waves should be regarded as fast-mode magnetosonic waves in nature. 

Previous statistical studies have revealed that EUV waves are excited by CMEs rather that flare pressure pulses \citep[e.g.,][]{2002ApJ...572L..99C,2006ApJ...641L.153C,2011ApJ...732L..20C,2012ApJ...752L..23S,2017ApJ...851..101S,2017SoPh..292....7L,2013AA...556A.152X}. However, the eruptions studied in the present paper did not cause any CME in coronagraphs, which suggests that the observed EUV waves were driven by other physical processes rather than CMEs. Based on the observational results, we propose that the observed EUV waves on the eastern side of the transequatorial loop systems in the present two cases were driven by the sudden lateral expansion of the loop systems due to the impingement of the associated coronal jets. In addition, the observed QFP waves along the loop system that connected AR11159 and AR11161 in the first event was formed by the dispersive evolution of the disturbance caused by the interaction between the EUV wave and the loop system \citep{2018MNRAS.477L...6S}.

In the driving mechanism of EUV waves proposed in the present paper, the sudden change of magnetic and plasma pressures perpendicular to the transequatorial loop systems may need to be fast enough to drive EUV waves. The fast moving coronal jets, that possess high kinetic energy, provide enough momentum to drive the lateral expansion of the loop systems and therefore increase the magnetic and plasma pressures. The lifetimes of the EUV waves in the present two events were all less than 10 minutes, which are much shorter than the typically hour-long lifetime of EUV waves \citep{2014SoPh..289.3233L}. This is possibly because of that the driven times of the EUV waves due to sudden lateral expansion of the transequatorial loop systems are much shorter than those driven by large-scale CMEs. Obviously, CMEs can provide a continuous driving force for EUV waves during the initial stage, whereas the lateral expansion of loops caused by coronal jets can only drive EUV waves within a short time interval. Since the ejection of coronal jets are transient phenomena, the expanded loop systems will return to their initial equilibrium positions when the ejection of the coronal jets finished. Therefore, the short lifetimes of the observed EUV waves also support the scenario that the observed EUV waves were driven by the sudden lateral expansion of the transequatorial loop systems due to the interaction of coronal jets. \cite{2015ApJ...804...88S} also reported an EUV wave that was excited by the expansion motion of coronal loops. We noted that the lifetime of the EUV wave studied in \cite{2015ApJ...804...88S} was about 6 minutes, which is in agreement with our result that EUV waves driven by loop expansions should have a shorter lifetimes than those driven by CMEs.

The formation mechanisms of the loop expansions in \cite{2015ApJ...804...88S} and the present cases are totally different. In \cite{2015ApJ...804...88S} the loop expansion was caused by the expansion of the newly formed reconnected loops resembling of the mechanism of a slingshot in the eruption of coronal jets. In the present study, the expansion of the loop systems were directly cased by the impingement of fast  coronal jets. Although the different formation mechanisms of the coronal loop expansion, the driving mechanism of the EUV waves in these cases should be similar to each other; namely, all of them were driven by the fast expanding coronal loops. In the future, more similar observational and theoretical studies are required to test the scenario proposed in the present paper.

\acknowledgments
We thank the excellent observations provided by the {\em SDO} team, and the referee's valuable comments that largely improved the quality of the paper. This work is supported by the Natural Science Foundation of China (11773068, 11633008, 11403097, 11503084,11773038), the Yunnan Science Foundation (2015FB191,2017FB006), the Specialized Research Fund for State Key Laboratories, the Open Research Program of CAS Key Laboratory of Solar Activity (KLSA201813), and the Youth Innovation Promotion Association (2014047) of Chinese Academy of Sciences.

\end{document}